\documentclass[conference,final,10pt]{IEEEtran}
\usepackage[a4paper, left=1.2cm, right=1.2cm, bottom=2.5cm, top=1.8cm]{geometry}
\usepackage[utf8]{inputenc}
\usepackage{graphicx}
\usepackage{amsmath,amssymb,amsthm,mathtools}
\usepackage{color}
\usepackage[linesnumbered,ruled]{algorithm2e}
\usepackage{upgreek}
\usepackage{ragged2e}
\usepackage{stfloats}
\usepackage{float}
\usepackage{bm}
\usepackage{cite}
\usepackage{pmat}
\usepackage{cases}
\usepackage{subfigure}
\usepackage{multicol,lipsum}

\usepackage[acronym,shortcuts]{glossaries}

\newacronym{5G}{5G}{fifth generation}
\newacronym{BS}{BS}{base station}
\newacronym{MIMO}{MIMO}{Multiple-Input Multiple-Output}
\newacronym{MISO}{MISO}{Multiple-Input Single-Output}
\newacronym{SISO}{SISO}{Single-Input Single-Output}
\newacronym{RX}{RX}{receiver}
\newacronym{TX}{TX}{transmitter}
\newacronym{BF}{BF}{beamforming}
\newacronym{SotA}{SotA}{State-of-the-Art}
\newacronym{NOMA}{NOMA}{Non-orthogonal Multiple Access}
\newacronym{PD-NOMA}{PD-NOMA}{Power Domain NOMA}
\newacronym{AWGN}{AWGN}{additive white Gaussian noise}
\newacronym{SINR}{SINR}{signal-to-interference-plus-noise ratio}
\newacronym{FP}{FP}{fractional programming}
\newacronym{SDR}{SDR}{semidefinite relaxation}
\newacronym{SDP}{SDP}{semidefinite programming}
\newacronym{CSI}{CSI}{channel state information}
\newacronym{OMA}{OMA}{Orthogonal Multiple Access}
\newacronym{TDMA}{TDMA}{Time Division Multiple Access}
\newacronym{CDMA}{CDMA}{Code Division Multiple Access}
\newacronym{OFDMA}{OFDMA}{Orthogonal Frequency Division Multiple Access}
\newacronym{mMTC}{mMTC}{massive Machine Type Communications}
\newacronym{DoF}{DoF}{Degree of Freedom}
\newacronym{CB}{CB}{Cluster-based}
\newacronym{BB}{BB}{Beamforming-based}
\newacronym{SIC}{SIC}{Successive Interference Cancellation}
\newacronym{QoS}{QoS}{Quality of Service}
\newacronym{i.i.d}{i.i.d}{independent and identically distributed}
\newacronym{PDF}{PDF}{probability density function}
\newacronym{CDF}{CDF}{cumulative density function}

\renewcommand{\smallskip}{\vspace{0.25cm}}

\newcommand{\abs}[1]{\left|{#1}\right|}

\newcommand{\tr}[1]{{\rm Tr}\left({#1}\right)}
\newcommand{\norm}[1]{\left\|{#1}\right\|_{2}}

\newcommand{\referencesrootdir}{./}
\newcommand{\myreferences}{\referencesrootdir/listofpublications.bib}

\begin{document}

\title{Fractional Programming for Robust TX BF Design in Multi-User/Single-Carrier PD-NOMA\\[-1ex]}

\author{
\IEEEauthorblockN{Hiroki Iimori$^{* \dagger}$, Giuseppe Abreu$^{\dagger}$ and Koji Ishibashi$^{*}$}
\IEEEauthorblockA{
$^{*}$ Advanced Wireless \& Communication Research Center (AWCC), The University of Electro-Communications\\
1-5-1 Chofugaoka, Chofu-shi, Tokyo 182-8585, Japan\\
{\tt h.iimori@ieee.org, koji@ieee.org}\\
$^\dagger$ Department of Computer Science and Electrical Engineering, Jacobs University Bremen\\ Campus Ring 1, 28759, Bremen, Germany \\
{\tt g.abreu@jacobs-university.de}\\[-1.5ex]
}
}
\maketitle

\begin{abstract}
We present a new \ac{BB} \ac{MISO}-\ac{NOMA} scheme for \ac{PD-NOMA}, in which the total transmit power consumption is minimized subjected to prescribed \ac{SINR} requirements for each user, and under the assumption that only imperfect \ac{CSI} is available at the transmitter.
To this end, the \ac{FP}-based quadratic transform is employed to reformulate the non-convex \ac{SINR} constraint of the original problem into a tractable quadratic form, which contains an estimate of the \ac{CSI} error vector as a parameter.
Taking advantage of the fact that the zero duality gap holds for the non-convex quadratic problems, a closed-form expression for an estimate of the \ac{CSI} error vector is derived, completing the formulation.
Finally, a novel iterative algorithm based on both the herein derived \ac{CSI} error vector and the \ac{SDR} technique is contributed, which is shown to capable of efficiently solving the constrained min-power problem.
Simulation results are given which illustrate the effectiveness of the proposed algorithm, which is found to sacrifice only small quantities of transmit power in return for substantial increase in robustness against \ac{CSI} imperfection. 
\end{abstract}

%\begin{IEEEkeywords}
%
%\end{IEEEkeywords}

% Introduction ------------------------------------
\section{Introduction}
\label{sect:intro}
\vspace{-0.5ex}

\emph{Non-orthogonality} is a key concept in \ac{5G} and beyond networks, which aims to improve the efficiency of resource utilization via the carefully designed overlapping of wireless signals.
In particular, the advantages of code and power domain \acf{NOMA} over classic \ac{OMA} approaches such as \ac{TDMA} \cite{Glisic1997}, \ac{CDMA} \cite{Viterbi1995} and \ac{OFDMA} \cite{Holma2009}, have been demonstrated in \cite{Ding2017} and references thereby.
The \ac{NOMA} approach has since become the multiplexing method of consensus  to ensure massive connectivity in future wireless networks, while resulting also in significant enhancement of spectrum efficiency \cite{Saito2013, Ding2017, LiuPIEEE2017, }, being incorporated in the ongoing standardization of \ac{5G} radio, Release 14 and beyond, which determined that non-orthogonal transmission should be considered at least in the uplink of \ac{5G} \ac{mMTC} systems.  

In turn, the combination of \ac{MIMO} techniques and \ac{NOMA} has attracted  significant attention from both Academia and Industry, as it enables further improvement thanks to the expansion of \acp{DoF} in spatial domain.

Two distinct strategies for \ac{MIMO}-\ac{NOMA} system designs can be found in the literature, namely, a) the \ac{CB} \ac{MIMO}-\ac{NOMA} \cite{DingTWC2016,Elkashlan2016,AliAccess2016} and b) the \acf{BB} \ac{MIMO}-\ac{NOMA} \cite{HanifTSP2016,ChoiTWC2016,AlaviTVT2018} designs.
In the \ac{CB} \ac{MIMO}-\ac{NOMA} approach, users in a cell are partitioned into groups (clusters), with a conventional \ac{SIC}-based \ac{PD-NOMA} scheme applied within each group, while transmit beamforming vectors are designed for each cluster.
The advantage of this scheme is that since the additional spatial \ac{DoF} is devoted to ensuring that the appropriately designed  beam corresponding to a specific cluster holds orthogonality to users assigned to other clusters, the inter-cluster interference can be sufficiently mitigated.
On the other hand, the drawback of the \ac{CB} \ac{MIMO}-\ac{NOMA} approach is the combinatorial optimization problem which needs to be solved in order to form the clusters, which becomes prohibitive (NP-hard) for large number of users.

In contrast, in \ac{BB} \ac{MIMO}-\ac{NOMA}, different transmit beamforming vectors are assigned to different users individually, which then perform \ac{SIC} in an order defined by the relative channel gains associated with each user.
In this approach, the computational cost at the \ac{BS} is significantly reduced due to the fact that user grouping is not required, such that \ac{BB} \ac{MIMO}-\ac{NOMA} is more scalable than its counterpart.
However, due to the larger number of users whose interference need be mitigated, it is easy to foresee that \ac{BB} \ac{MIMO}-\ac{NOMA} is more sensitive to the accuracy of instantaneous \acf{CSI} knowledge than \ac{CB} \ac{MIMO}-\ac{NOMA}.

Interestingly, despite the great amount of effort dedicated thus far to develop \ac{NOMA} technologies, not enough attention has been payed to the aforementioned problem as evidence by the fact that perfect \ac{CSI} knowledge at the transmitter is the standard assumption in related literature.
While usually a reasonable simplification in wireless communications systems, the assumption of perfect \ac{CSI} is particularly unrealistic in \ac{NOMA}  due to inherent overloading, which is known to impact the accuracy of channel estimation \cite{MyListOfPapers:StanzakTSP2006}.

In this paper, we therefore turn our attention to the design of a multi-user/single-carrier \ac{BB} \ac{MISO}-\ac{NOMA} system subjected to norm-bounded channel imperfection, proposing a novel \acf{FP}-based transmit beamforming design aiming at the minimization of the total power consumption.

The remainder of the article is as follows.
Section \ref{System_Model} describes the system model and the problem formulation aiming at minimizing the total transmit power subjected to \acf{SINR} requirements for each user.
The \ac{QoS} based min-min problem is reformulated and solved in Section \ref{Sec:Proposed}. The mathematical expression of the \ac{CSI} error vector estimate is derived therein, and the proposed algorithm are also offered thereby.
Simulation results illustrating the effectiveness of the proposed algorithm compared with the non-robust scheme -- that is, subject to the same channels, but optimized under the assumption of perfect \ac{CSI} -- are shown in Section \ref{Sec:SimulationResults}.
Conclusions and discussions on possible future works are given in Section \ref{Sec:Conclusion}.

\subsection{Notation}

Throughout the article, matrices and vectors are expressed respectively by bold capital and small letters, such as in $\bm{X}$ and $\bm{x}$.
The conjugate, Hermitian (transpose conjugate) and inverse operators are be respectively denoted by $\left(\cdot\right)^{*}$, $\left(\cdot\right)^\mathrm{H}$ and $\left(\cdot\right)^{-1}$, while the $\ell_{2}$-norm operators are be depicted by $\norm{\cdot}$.
A complex matrix with $a$ columns and $b$ rows is denoted by $\bm{X}\in\mathbb{C}^{a\times b}$, and a circularly symmetric complex random scalar variable following the complex Gaussian distribution with mean $\mu$ and variance $\sigma^2$ is expressed as $x \sim \mathcal{CN}\left(\mu,\sigma^2\right)$.
For given $\mathbf{A}, \mathbf{B}\in\mathcal{S}^{n}$, $\mathbf{A}\succeq\mathbf{B}$ indicate that $\mathbf{A}-\mathbf{B}$ is positive semidefinite, and
${\rm Re}(\cdot)$ and ${\rm R}(\mathbf{A})$ express the real part of a complex number and the range space of the matrix $\mathbf{A}$, respectively.

% System Model ------------------------------------

\section{System Model}
\label{System_Model}

Consider a power domain single-carrier \ac{NOMA} system where one \ac{BS}, equipped with $N_{t}$ transmit antennas and capable of performing digital precoding towards all users, attempts to simultaneously transmit $U$ data-stream to a pool of single-antenna downlink users $u\in\{1,2,\ldots,U\}$. 

Without loss of generality, let all channel vectors $\mathbf{h}_{u}$ between the \ac{BS} and the $u$-th user be sorted in ascending order, such that $\norm{\mathbf{h}_{1}}\leq\norm{\mathbf{h}_{2}}\leq\ldots\leq\norm{\mathbf{h}_{U}}$.
It is assumed then that each $u$-th user in such a system can decode its own intended signal $s_u$ after successively detecting and removing the first $u-1$ users's messages while regarding the higher order signals as interference.
It follows from this assumption that, for a given reference user $u$, all {{$\ell_u$}} users with ${{\ell_u}}\in\{u,u+1,\ldots,U\}$ ought to detect the $u$-th user's signal in order to successfully decode their own intended signals.

Letting $\mathbf{w}_{u} \in\mathbb{C}^{N_{t}\times 1}$ denote the transmit beamforming vector corresponding to the symbol intended for the $u$-th user, the received signal at the ${{\ell_u}}$-th expressed so as to highlight the $u$-th user's signal can be written as
\begin{equation}\label{eqn:receivedsignal}
  y_{u,\ell_u} = 
  \mathbf{h}^{\rm H}_{\ell_u} \mathbf{w}_{u} s_{u}
  + \sum_{m=1}^{u-1}\mathbf{e}^{\rm H}_{\ell_u}  \mathbf{w}_m  s_m 
  + \sum_{k=u+1}^{U}\mathbf{h}^{\rm H}_{\ell_u}  \mathbf{w}_k  s_k 
  + n_{\ell_u},
\end{equation}
where $n_{\ell}\sim \mathcal{CN}(0,\sigma^2)$ is the circularly symmetric \ac{AWGN} at user {$\ell_u$} and 
\begin{equation}
\mathbf{h}_{\ell_u} = \hat{\mathbf{h}}_{\ell_u} +\mathbf{e}_{\ell_u}, \forall \ell_u
\end{equation}
with {$\hat{\mathbf{h}}_{\ell_u}\in\mathbb{C}^{M\times 1}$ and $\mathbf{e}_{\ell_u}\in\mathbb{C}^{M\times 1}$} denoting the estimate of the physical channel {$\mathbf{h}_{\ell_u}$} between the \ac{BS} and the user {$\ell_u$} and the corresponding channel estimation error bounded such that {$\norm{\mathbf{e}_{\ell_u}}\leq \varepsilon$}, respectively.

Let {$\text{SINR}_{u,\ell_u}$ } denote the \ac{SINR} of the signal $s_u$ at the user {$\ell_u$}.
Then, taking into account the fact that the achievable throughput of the $u$-th user is determined by the lowest {$\text{SINR}_{u,\ell_u}$ } among all {$\ell_u$} users, the maximal data rate of the $u$-th user in order for all users to be able to successfully obtain their own intended signals is given by
\begin{equation}\label{eqn:Throughput}
R_u=\log_2 \big(1+\min_{\ell_u} \text{SINR}_{u,\ell_u}\big),
\end{equation}

\begin{figure}[H]
\center
\includegraphics[width=\columnwidth]{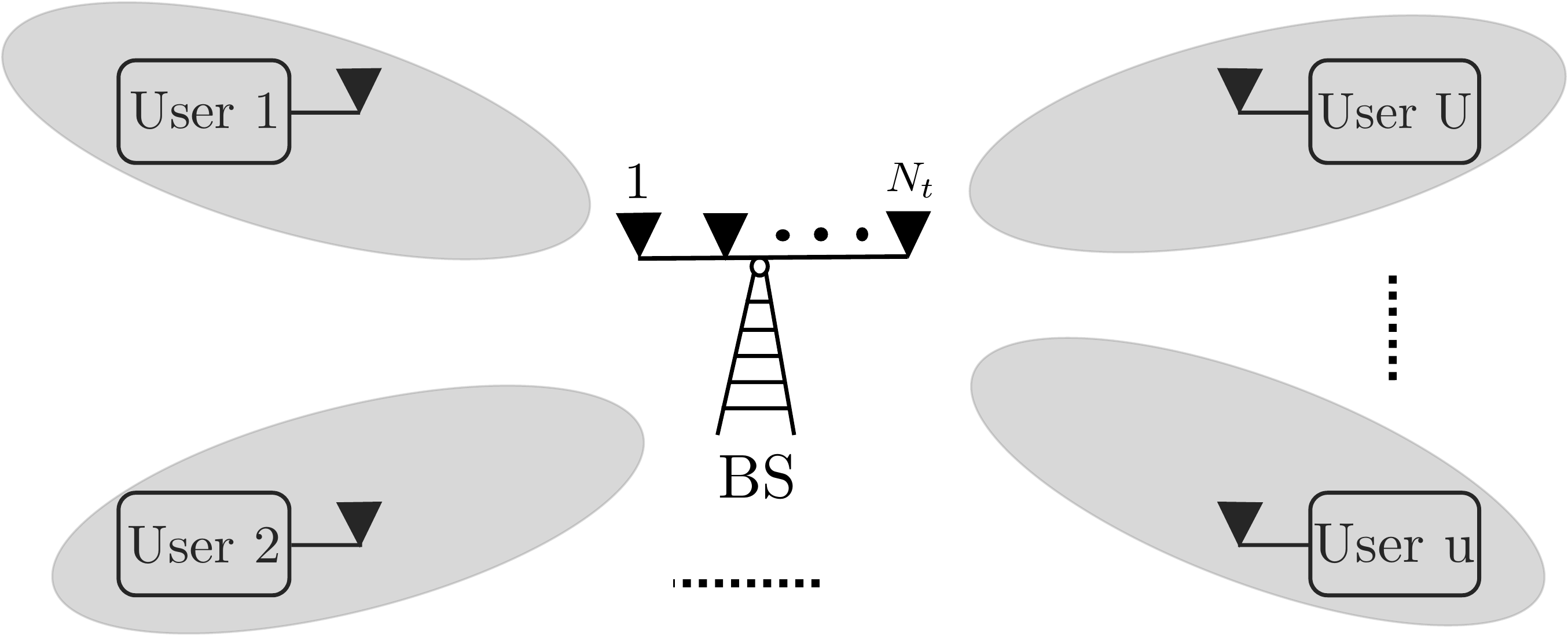}
\caption[]{System Model of \ac{BB} \ac{MISO} \ac{NOMA} with $N_t$ transmit antennas and $U$ single-antenna users.}
\label{fig:System_model}
\vspace{-1ex}
\end{figure}

\noindent with
{
\begin{equation}
\label{eqn:SINR}
 \text{SINR}_{u,\ell_u}= \frac{\mathbf{h}^{\rm H}_{\ell_u}\mathbf{w}_{u}\mathbf{w}^{\rm H}_{u}\mathbf{h}_{\ell_u}}
 {{\displaystyle \sum_{m=1}^{u-1}}\mathbf{e}^{\rm H}_{\ell_u}\mathbf{w}_m\mathbf{w}^{\rm H}_m\mathbf{e}_{\ell_u}
 + {\displaystyle \sum_{k=u+1}^{U}}\mathbf{h}^{\rm H}_{\ell_u}  \mathbf{w}_k\mathbf{w}^{\rm H}_k\mathbf{h}_{\ell_u}
 +\sigma^2}.
\end{equation}\vspace{-1ex}
}

Although several \ac{BF} techniques aiming at maximizing the downlink sum data rate in \ac{PD-NOMA} systems have been proposed in the past few years \cite{HanifTSP2016}, we argue that the sum rate criterion in an overkill, considering that it is sufficient to maintain a certain \ac{SINR} in order for a user to enjoy its intended application.
We therefore consider instead the optimization of transmit beamformers aimed at minimizing total power consumption minimization, while constrained to satisfying worst-case individual target throughput requirements, $i.e.$
\vspace{-1ex}
\begin{subequations}
\label{eqn:OP_Original1}
\begin{eqnarray}
\min_{\mathbf{w}_{u}, \forall u}&& \sum^{U}_{u=1}\norm{\mathbf{w}_{u}}^2\\
{\rm s.t.}&& \min_{\norm{\mathbf{e}_{\ell_u}}\leq\varepsilon} R_{u} \geq \log_2(1 + \Gamma_{u})\:\:\forall u,
\end{eqnarray}
\end{subequations}
where $\Gamma_{u}$ denotes the target SINR for the $u$-th user, and $\mathbf{e}_{\ell_u}$ is the worst-case \ac{CSI} error vector corresponding to user {$\ell_u$}.

From the fact that the logarithmic function is a non-decreasing function, the above optimization problem can be readily rewritten as 
\begin{subequations}
\label{eqn:OP_Original2}
\begin{eqnarray}
\min_{\mathbf{w}_{u}, \forall u}&& \sum^{U}_{u=1}\norm{\mathbf{w}_{u}}^2\\
{\rm s.t.}&& \min_{\norm{\mathbf{e}_{\ell_u}}\leq\varepsilon}\min_{\ell_u}\: \text{SINR}_{u,\ell_u}\geq \Gamma_{u}\:\: \forall u. 
\label{eqn:SINRConstraint}
\end{eqnarray}
\end{subequations}

It can be recognized from the constraint in inequality \eqref{eqn:SINRConstraint}, that the problem formulated in equation \eqref{eqn:OP_Original2} is not convex and therefore intractable in its original form.
In the next section, we therefore tackle the above problem by 1) transforming the $\min \min$ operation in \eqref{eqn:SINRConstraint}; and 2) recasting the non-convex ratio constraints into a tractable quadratic convex form.

\section{Proposed robust beamforming design}
\label{Sec:Proposed}

\subsection[]{Transformation of \ac{SINR} Constraints}
\label{sec:ConstRelaxation}

Following related literature \cite{AlaviCL2017}, each constraint on $\Gamma_{u}$ constructed with basis on the $\min$ operator over all {$\ell_u$}s in equation \eqref{eqn:SINRConstraint} can be replaced by a set of $(U-u+1)$ simultaneous constraints.

That is, $\forall u$,
{
\begin{eqnarray}
\label{eqn:RelaxedSINR1}
\min_{\norm{\mathbf{e}_{\ell_u}}\leq\varepsilon}\min_{\ell_u}\: \text{SINR}_{u,\ell_u}\geq \Gamma_{u} \!\equiv
\!\left\{\!\!\!\!\!
\begin{array}{c}
{\displaystyle\min_{\norm{\mathbf{e}_{\ell_u}}\leq\varepsilon}}\: \text{SINR}_{u,u} \geq \Gamma_{u}, \\
{\displaystyle\min_{\norm{\mathbf{e}_{\ell_u}}\leq\varepsilon}}\: \text{SINR}_{u,u+1} \geq \Gamma_{u},\\
\vdots\\[1ex]
{\displaystyle\min_{\norm{\mathbf{e}_{\ell_u}}\leq\varepsilon}}\: \text{SINR}_{u,U} \geq \Gamma_{u}.
\end{array}
\right.\hspace{-5ex}
\end{eqnarray}}

\vspace{-1ex}
Thanks to the recasted constraints above, the original optimization problem given in equation \eqref{eqn:OP_Original2} can be rewritten as
\vspace{-1ex}
\begin{subequations}
\label{eqn:OP_transformed1}
\begin{eqnarray}
\min_{\mathbf{w}_{u}, \forall u}&& \sum^{U}_{u=1}\norm{\mathbf{w}_{u}}^2\\
{\rm s.t.}&& \min_{\norm{\mathbf{e}_{\ell_u}}\leq\varepsilon}\: \text{SINR}_{u,\ell_u}\geq \Gamma_{u}\:\: \forall u, \ell_u. 
\label{eqn:SINRConstraint_transformed1}
\end{eqnarray}
\end{subequations}
where, for clarity, we highlight that the constraints in \eqref{eqn:SINRConstraint_transformed1} are for all $u$ and {$\ell_u$}.

\subsection{Quadratic Transform of Ratio Constraints}
\label{sec:ProposedMain}

Unfortunately, the problem formulated in equation \eqref{eqn:OP_transformed1} is still not convex, due to the SINRs defined in equation \eqref{eqn:SINR}.
While several methods such as the the Taylor series approximation \cite{Bjornson2013} and the \acf{SDR} \cite{AlaviCL2017} have been proposed for the transformation of non-convex ratios constraints in the past decade, a novel quadratic transformation technique for such non-convex ratio problems has been recently proposed in \cite{ShenTSP2018}, which has been shown not to result in any approximation gap at the optimal point.

Indeed, consider a generic maximization problem with a sum of ratios as objective, such as
\begin{subequations}
\begin{eqnarray}
\label{eqn:GeneralFP}
\max_{\bm{x}}&& \sum^{M}_{m=1}\bm{a}^\mathrm{H}_{m}\left(\bm{x}\right)\mathbf{B}^{-1}_{m}\left(\bm{x}\right)\bm{a}_{m}\left(\bm{x}\right)\\
{\rm s.t.}&& \bm{x}\in \mathcal{X}, 
\end{eqnarray}
\end{subequations}
where $\bm{a}_{m}\left(\bm{x}\right)$ denotes an arbitrary complex function, $\mathbf{B}_{m}\left(\bm{x}\right)$ is an arbitrary symmetric positive definite matrix, and $\bm{x}$ is a variable to be optimized in a constraint set $\mathcal{X}$.

Then, the equivalent problem after applying the \emph{quadratic transformation}  \cite{ShenTSP2018, IimoriTWC19, IimoriWCNC19} can be written as 
\begin{subequations}
\begin{eqnarray}
\label{eqn:GeneralFP_QT}
\max_{\bm{x}}&& \sum^{M}_{m=1}2\mathrm{Re}\left\{\mathbf{t}^\mathrm{H}_{m}\bm{a}_{m}\left(\bm{x}\right)\right\} - \mathbf{t}^\mathrm{H}_{m}\mathbf{B}_{m}\left(\bm{x}\right)\mathbf{t}_{m}\\
{\rm s.t.}&& \bm{x}\in \mathcal{X},\:\: \mathbf{t}_{m}\in\mathbb{C},
\end{eqnarray}
\end{subequations}
where $\mathbf{t}_{m}$ is a scaling quantity designed so as to ensure that, for a given $\bm{x}$,  the original function in equation \eqref{eqn:GeneralFP} is equivalent to the transformed function as outlined in equation \eqref{eqn:GeneralFP_QT} and given by
\begin{equation}
\mathbf{t}_{m} = \mathbf{B}^{-1}_{m}\left(\bm{x}\right)\bm{a}_{m}\left(\bm{x}\right).
\end{equation}

From the above, equation \eqref{eqn:SINR} can be rewritten in the form
\vspace{-1ex}
{
\begin{eqnarray}
\label{eqn:SINRFPTransformed}
 \text{SINR}_{u,\ell_u} \!\!\!\!\!\!\!&=&\!\!\!\!\!\! 2\mathrm{Re}\{t^{*}_{u,\ell_u}\mathbf{h}^{\rm H}_{\ell_u} \mathbf{w}_{u}\}\\
 \!\!\!\!\!\!\!\!&&\!\!\!\!\!\!- |t_{u,\ell_u}\!|^2\!\Bigg[\!{\displaystyle \sum_{m=1}^{u-1}}\!\mathbf{e}^{\rm H}_{\ell_u}\!\mathbf{w}_m\!\mathbf{w}^{\rm H}_m\mathbf{e}_{\ell_u} 
 \!\!+\!\!\!\!\! {\displaystyle \sum_{k=u+1}^{U}}\!\!\!\mathbf{h}^{\rm H}_{\ell_u} \! \mathbf{w}_k\!\mathbf{w}^{\rm H}_k\mathbf{h}_{\ell_u}
 \!\!\!+\!\sigma^{2}\!\Bigg]\!,\nonumber
\end{eqnarray}}
with
{\begin{equation}\label{eqn:t}
t_{u,\ell_u}\!\!=\!\!\Bigg[{\displaystyle \sum_{m=1}^{u-1}}\!\mathbf{e}^{\rm H}_{\ell_u}\!\mathbf{w}_m\!\mathbf{w}^{\rm H}_m\mathbf{e}_{\ell_u} 
 \!\!+\!\!\!\! {\displaystyle \sum_{k=u+1}^{U}}\!\!\!\mathbf{h}^{\rm H}_{\ell_u} \! \mathbf{w}_k\!\mathbf{w}^{\rm H}_k\mathbf{h}_{\ell_u}
 \!\!+\sigma^{2}\!\Bigg]^{\!-1}\!\!\!\!\!\!\mathbf{h}^{\rm H}_{\ell} \mathbf{w}_{u}.
 \end{equation}}

\vspace{-2.5ex}
\subsection{Closed-form of Worst Sum-SINR CSI Error Vector}
\vspace{-0.5ex}

With possession of equation \eqref{eqn:SINRFPTransformed}, the optimization problem in equation \eqref{eqn:OP_transformed1} can be recast as a quadratically constrained quadratic convex problem with respect to $\mathbf{w}_{u}$.
To this end, however, all {$\text{SINR}_{u,\ell_u}$}'s in the inequalities \eqref{eqn:SINRFPTransformed} must be optimized with respect to the \ac{CSI} error vector {$\mathbf{e}_{\ell_u}$}, which can lead to overwhelming complexity due to the large number of distinct pairs {$u,\ell_u$}.
In order to keep overall complexity under control, {{we introduce a new variable $\ell\in\{1,2,\ldots,U\}$}} and circumvent this problem by relaxing these requirement into the sum-\ac{SINR} minimization problems (for each {$\ell$})
\vspace{-1ex}
\begin{subequations}\label{eqn:SumSINRMin}
\begin{eqnarray}
\min_{\mathbf{e}_{\ell}}&& \sum^{\ell}_{j=1}\text{SINR}_{j,\ell}\\
{\rm s.t.}&& \norm{\mathbf{e}_{\ell}}\leq\varepsilon,
\end{eqnarray}
\end{subequations}
which under the quadratic transform yields
\begin{subequations}
\label{eqn:InnerOP}
\begin{eqnarray}
\min_{\mathbf{e}_{\ell}}&& -\mathbf{e}^{\rm H}_{\ell}\mathbf{A}_{\ell}\mathbf{e}_{\ell}+2\mathrm{Re}\{\mathbf{e}^{\rm H}_{\ell}\mathbf{b}_{\ell}\}+ c_{\ell}\\
{\rm s.t.}&& \norm{\mathbf{e}_{\ell}}\leq\varepsilon,
\label{eqn:Inner_Const}
\end{eqnarray}\vspace{-1ex}
\end{subequations}
\noindent where
{
\begin{subequations}
\begin{eqnarray}
\mathbf{A}_{\ell} &=&\sum^{\ell}_{j=1}\abs{t_{j,\ell}}^2\sum_{i\neq j}\mathbf{w}_{i}\mathbf{w}^{\rm H}_{i},\\
\mathbf{b}_\ell &=& \sum^{\ell}_{j=1}t^{*}_{j,\ell}\mathbf{w}_j - \abs{t_{j,\ell}}^2\sum^{U}_{k=j+1}\mathbf{w}_{k}\mathbf{w}^{\rm H}_{k}\hat{\mathbf{h}}_{\ell},\\
c_\ell &=& \sum^{\ell}_{j=1}2\mathrm{Re}\{t^{*}_{j,\ell}\hat{\mathbf{h}}^{\rm H}_{\ell}\mathbf{w}_j\}\\[-2ex]
&& \hspace{7ex}\!-\! \sum^{\ell}_{j=1}|t_{j,\ell}|^2\!\Bigg[ \sum^{U}_{k=j+1}\!\! \hat{\mathbf{h}}^{\rm H}_{\ell}  \mathbf{w}_k\mathbf{w}^{\rm H}_k\hat{\mathbf{h}}_{\ell} \!+\! \sigma^2\!\Bigg].\nonumber
\end{eqnarray}
\end{subequations}}\vspace{-1.5ex}

Notice that due to the relaxation of equation \eqref{eqn:RelaxedSINR1} into equation \eqref{eqn:SumSINRMin}, the \ac{CSI} error vector $\mathbf{e}_{\ell}$ obtained is a worst-case vector in the sum-SINR sense.
Furthermore, since $\mathbf{A}_{\ell}\succeq 0$, the objective function in equation \eqref{eqn:InnerOP} is a concave function and therefore can not be minimized via numerical convex optimization tools.
Fortunately, the strong duality holds for non-convex quadratic problems \cite{Zheng2012, Pong2014}, so that equation \eqref{eqn:InnerOP} can be solved via techniques such as \ac{SDR} and the Lagrange multiplier method.

The Lagrangian function of equation \eqref{eqn:InnerOP} is given by
\begin{equation}
\mathcal{L}(\mathbf{e}_\ell,\lambda_{\ell}) = \mathbf{e}^{\rm H}_{\ell}\big(\lambda_{\ell}\mathbf{I}-\mathbf{A}_{\ell}\big)\mathbf{e}_{\ell}+2\mathrm{Re}\{\mathbf{e}^{\rm H}_{\ell}\mathbf{b}_{\ell}\}+ c_{\ell} - \lambda_{\ell}\varepsilon^2
\end{equation}
where $\lambda_{\ell}\geq0$ denotes the dual variable.

For a fixed $\lambda_{\ell}$, the optimal \ac{CSI} error vector $\mathbf{e}_{\ell}$ and the corresponding dual function $g(\lambda_{\ell})$ are, respectively, given by \cite{IimoriSPAWC2018}
\begin{equation}\label{eqn:e_opt}
\mathbf{e}_{\ell} = - \big(\lambda_{\ell}\mathbf{I}-\mathbf{A}_{\ell}\big)^{-1}\mathbf{b}_{\ell},
\end{equation}
\vspace{-1ex}
and
\begin{equation}\label{eqn:dual}
g(\lambda_{\ell}) \!\!=\!\! \left\{\!\!\!\! \begin{array}{ll}
     - \mathbf{b}^{\rm H}_{\ell}\big(\lambda_{\ell}\mathbf{I}\!-\!\mathbf{A}_{\ell}\big)^{\!-1}\mathbf{b}_{\ell}  \!\!& \mathrm{if}\: \mathbf{b}_{\ell}\!\in\!\mathrm{R}(\lambda_{\ell}\mathbf{I}\!-\!\mathbf{A}_{\ell}) \\
     \hspace{15ex} + c_{\ell} \!-\! \lambda_{\ell}\varepsilon^2  \!\!&  \mathrm{and}\:\lambda_{\ell}\mathbf{I}-\mathbf{A}_{\ell} \succeq 0\\
   -\infty \!\!&  \text{(otherwise)}.
  \end{array}\right.
\end{equation}\vspace{-2ex}

By virtue of the Schur complement, the dual variable $\lambda_{\ell}$ is optimized by solving an epigraph form of the maximization of the dual function in equation \eqref{eqn:dual}, such that
\vspace{-1ex}
\begin{subequations}
\label{eqn:DualOP}
\begin{eqnarray}
\max_{\lambda_{\ell},\beta_{\ell}}&& \beta_{\ell}\\
\vspace{-1ex}
{\rm s.t.}&& \begin{bmatrix}
\lambda_{\ell}\mathbf{I}\!-\!\mathbf{A}_{\ell} & \mathbf{b}_{\ell}\\
\mathbf{b}^{\rm H}_{\ell} & c_{\ell} \!-\! \lambda_{\ell}\varepsilon^2 - \beta_{\ell}
\end{bmatrix} \succeq 0
\label{eqn:SDP_Const}\\
&& \lambda_{\ell} \geq 0.
\end{eqnarray}
\end{subequations}
\begin{proof}
See Appendix \ref{Appndx:A}.
\end{proof}\vspace{-2ex}

One can readily notice that equation \eqref{eqn:DualOP} is a standard \ac{SDP} problem with a linear objective function, which can be efficiently solved in polynomial time via interior point methods \cite{CVX2017}. 

\vspace{-1ex}
\subsection{Robust TX BF Algorithm}

With estimates of the \ac{CSI} error vectors $\mathbf{e}_\ell$ obtained as per equation \eqref{eqn:e_opt} in hand, the original optimization problem given in equation \eqref{eqn:OP_transformed1} becomes a standard non-convex power minimization problem, which can be relaxed and solved in a number of different manners.
One possibility is, for instance, to solve the problems via \ac{SDR}\cite{AlexSPM2010}.
In this case, defining {$\mathbf{W}_{u}=\mathbf{w}_{u}\mathbf{w}^{\rm H}_{u} \succeq 0$, $\mathbf{H}_{\ell_u}=\mathbf{h}_{\ell_u}\mathbf{h}^{\rm H}_{\ell_u} \succeq 0$ and $\mathbf{E}_{\ell_u}=\mathbf{e}_{\ell_u}\mathbf{e}^{\rm H}_{\ell_u} \succeq 0$}, equation \eqref{eqn:OP_transformed1} can be transformed into
\vspace{-1ex}
\begin{subequations}
\label{eqn:OP_SDR}
\begin{eqnarray}
\min_{\mathbf{W}_{u}, \forall u}&& \sum^{U}_{u=1}\tr{\mathbf{W}_{u}}\\[-1.5ex]
{\rm s.t.}&&  \tr{\mathbf{H}_{\ell_u}\mathbf{W}_{u}} - \Gamma_{u}\bigg[{\displaystyle \sum_{m=1}^{u-1}}\tr{\mathbf{E}_{\ell_u}\mathbf{W}_{m}} \\[-1.5ex]
&& \hspace{8ex} + {\displaystyle \sum_{k=u+1}^{U}}\tr{\mathbf{H}_{\ell_u}\mathbf{W}_{k}}\bigg]\geq \Gamma_{u}\sigma^2\:\: \forall u, \ell_u \nonumber\\
&& \mathbf{W}_{u} \succeq 0\:\: \forall u,
\end{eqnarray}
\end{subequations}
where we omitted the rank-one constraint.

We remark that although the relaxed problem in equation \eqref{eqn:OP_SDR} can be efficiently solved by interior point methods, it is not guaranteed that such relaxed solutions are rank-one.
It is known, however, that if the largest eigenvalue of $\mathbf{W}_{u}$ is sufficiently larger than the second, the approximation gap between the global optimum and the solution to equation \eqref{eqn:OP_SDR} is tight \cite{KaripidisTSP08, LuoSPM10}, while randomization procedures can be used to generate an approximate solution otherwise.

\begin{algorithm}[t]
\SetKwInOut{Input}{Input}
\Input{{Channel Estimate: $\hat{\mathbf{h}}_{\ell}\:\forall \ell$}\\
\hspace{0.1ex} Target \ac{SINR}: $\mathbf{\Gamma}_{u}\:\forall u$\\
\hspace{0.1ex} Maximum number of iteration: $i_\text{max}$\\
\hspace{0.1ex} Initial TX beamforming vector: $\mathbf{w}^{(0)}_{u}\:\forall u$\\
\hspace{0.1ex} \ac{CSI} error bounding parameter: $\varepsilon$}
Generate random \ac{CSI} error {$\mathbf{e}^{(0)}_{\ell}\:\forall \ell$ such that $\norm{\mathbf{e}_{\ell}}\leq\varepsilon$}.\\
Set $i = 0$.

\Repeat{$\delta<10^{-4}$ or reach the maximum iteration $i_\text{max}$}{
\noindent $i$ $\leftarrow$ $i+1$\\
{$t_{u,\ell_u}\:\forall u,\ell_u$ $\leftarrow$  Equation \eqref{eqn:t} for given $\mathbf{e}^{(i-1)}_{\ell},\mathbf{w}^{(i-1)}_{u}$}.\\
$\lambda_\ell \:\forall \ell$ $\leftarrow$ Solve \ac{SDP} in equation \eqref{eqn:DualOP}.\\
{$\mathbf{e}^{(i)}_{\ell}$ $\leftarrow$ Compute from equation \eqref{eqn:e_opt}.}\\
$\mathbf{w}^{(i)}_{u}$ $\leftarrow$ Solve SDP in equation \eqref{eqn:OP_SDR}.\\
Check convergence $\delta = \sum^{U}_{u=1}\norm{\mathbf{w}^{(i-1)}_{u} - \mathbf{w}^{(i)}_{u}}/(N_{t}\cdot U)$\\
}
\caption[]{\ac{FP}-based Robust TX BF for \ac{PD-NOMA}.}
\label{alg:Main}
\end{algorithm}
\setlength{\textfloatsep}{-3pt}

\vspace{-1ex}
\section{Simulation results}
\label{Sec:SimulationResults}

In this section, we evaluate via computer simulations the effectiveness of the proposed algorithm in a downlink \ac{BB} \ac{MIMO}-\ac{NOMA} system with $N_t = 3$ transmit antennas at the \ac{BS} serving $U = 3$ single-antenna users, comparing its performance against that of a ``Non-Robust'' scheme which does not take into account channel uncertainties, both under perfect and imperfect  \ac{CSI} knowledge conditions.

\begin{figure}[t!]
\includegraphics[width=\columnwidth]{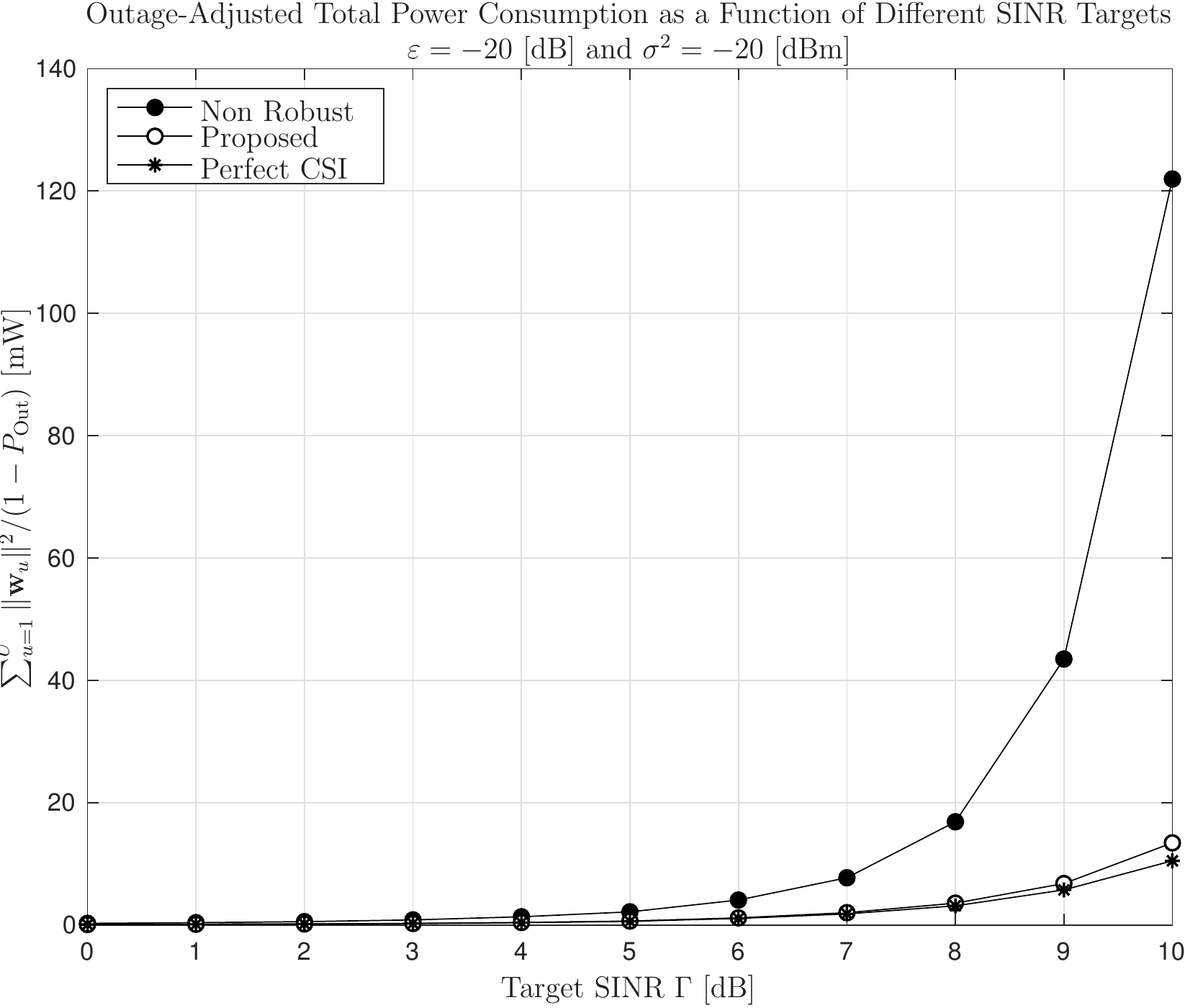}
\caption[]{Outage-adjusted total transmit power consumption at different SINR targets for imperfect \ac{CSI} error bounded according to $\varepsilon=-20$ [dB]}
\label{fig:PowervsSINRTarg}
\vspace{2ex}
\includegraphics[width=\columnwidth]{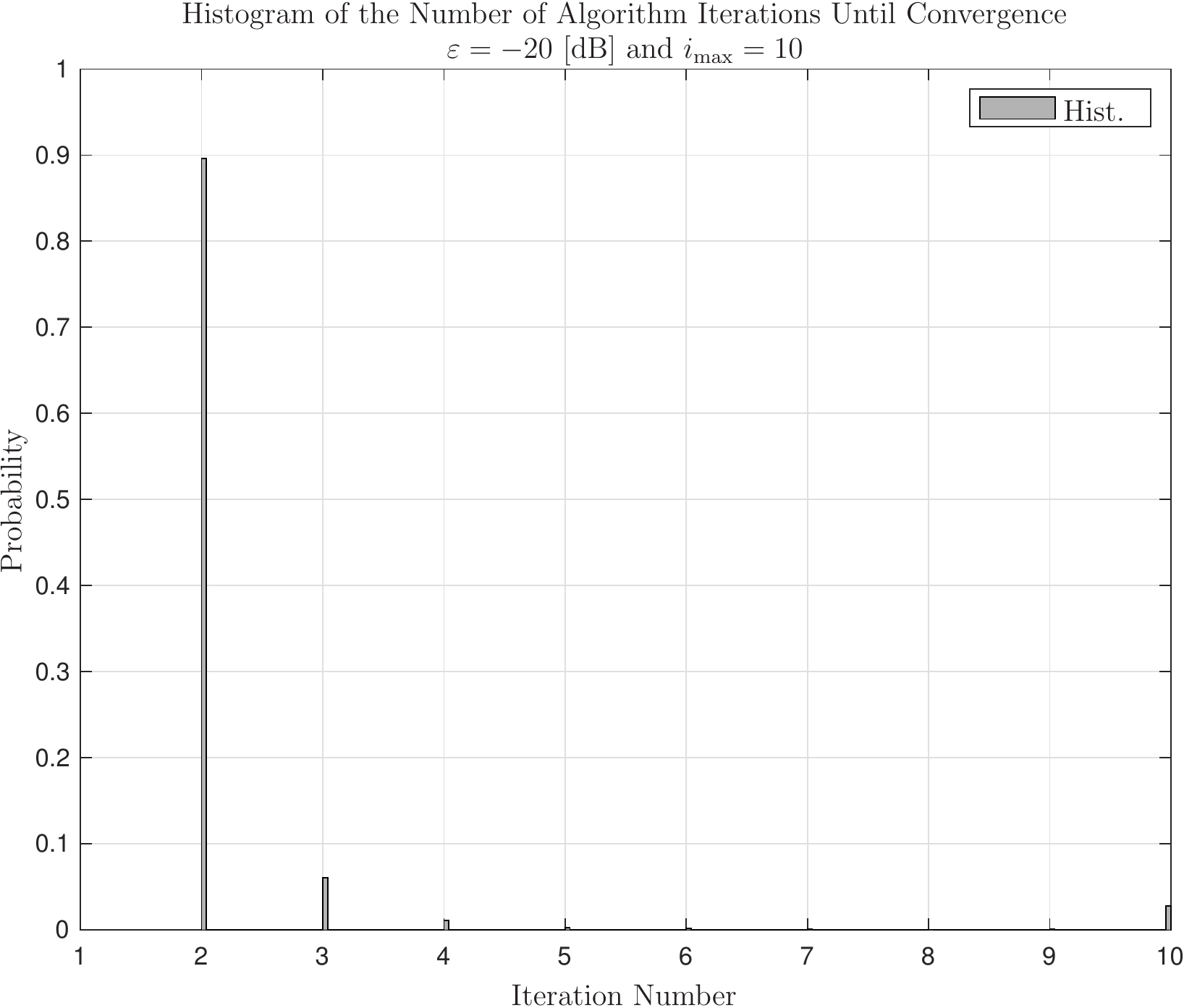}
\caption[]{Number of iterations until convergence of the proposed algorithm  with maximum number of iterations capped at $i_{\rm max}=10$.}
\label{fig:IterationHist}
\end{figure}

In all simulations, it is assumed that $\hat{\mathbf{h}}_{\ell}$ are \ac{i.i.d} Rayleigh fading channel estimates, with underlying vectors following a complex Gaussian distribution with zero-mean and variance adjusted by the number of transmit antennas $N_t$ for all $\ell$, namely, $\hat{\mathbf{h}}_{\ell}\sim\mathcal{CN}\big(\mathbf{0},\frac{1}{N_t}\mathbf{I}_{Nt}\big)$, while the \ac{CSI} error $\mathbf{e}_{\ell}$ is uniformly distributed within a disk of radius $0.01$ ($i.e.$, $\varepsilon=0.01$).
In addition, we use the following simulation setup unless otherwise mentioned.
The noise variance $\sigma^2$ is set to be $0.01$ [mW] and the number of iterations for the proposed algorithm is upper-bounded by $10$, $i.e.$, $i_{\rm max}=10$, whereas, for the sake of simplicity but without loss of generality, the target \ac{SINR} $\Gamma_{u}$ is assumed to be identical, namely, $\Gamma=\Gamma_{u}\:\forall u$, and scaled from $0$ [dB] to $10$ [dB].
All the results are averaged over $500$ channel realizations with $100$ \ac{CSI} error realizations for each channel realization.
Throughout the simulations, the MATLAB-based convex optimization toolbox CVX with its default solver SDPT3 is utilized to solve convex optimization problems such as \acp{SDP} in equation \eqref{eqn:DualOP} and equation \eqref{eqn:OP_SDR}.

First, in Figure \ref{fig:PowervsSINRTarg}, plots of the average transmit power consumption scaled by taking into account the outage probability defined by $P_{\rm Out}={\rm Pr}\left(\text{\ac{SINR}}<\Gamma\right)$ are shown, as a function of different target \acp{SINR}.
Figure \ref{fig:PowervsSINRTarg} clearly illustrates the consequence of ignoring \ac{CSI} imperfection, demonstrating that the proposed method is capable of significant power savings compared to the non-robust alternative, in fact coming very close to the performance of a system operating under perfect \ac{CSI}.

The latter results are even more motivating when considered next to the results  offered in Figure \ref{fig:IterationHist}, which shows the empirical probability mass function of the number of iterations till convergence required by for the proposed algorithm.
The figure indicates that in approximately $90$\% of the times convergence is achieved after only $2$ iterations, and the maximum allowed number of iterations ($i_{\rm max}=10$) is required less than $5$\% of the times.
\vspace{-2ex}
\begin{figure}[b!]
%\centering
\subfigure[Empirical PDF.]
{\label{fig:SINRHistat0dB}
\includegraphics[width=\columnwidth]{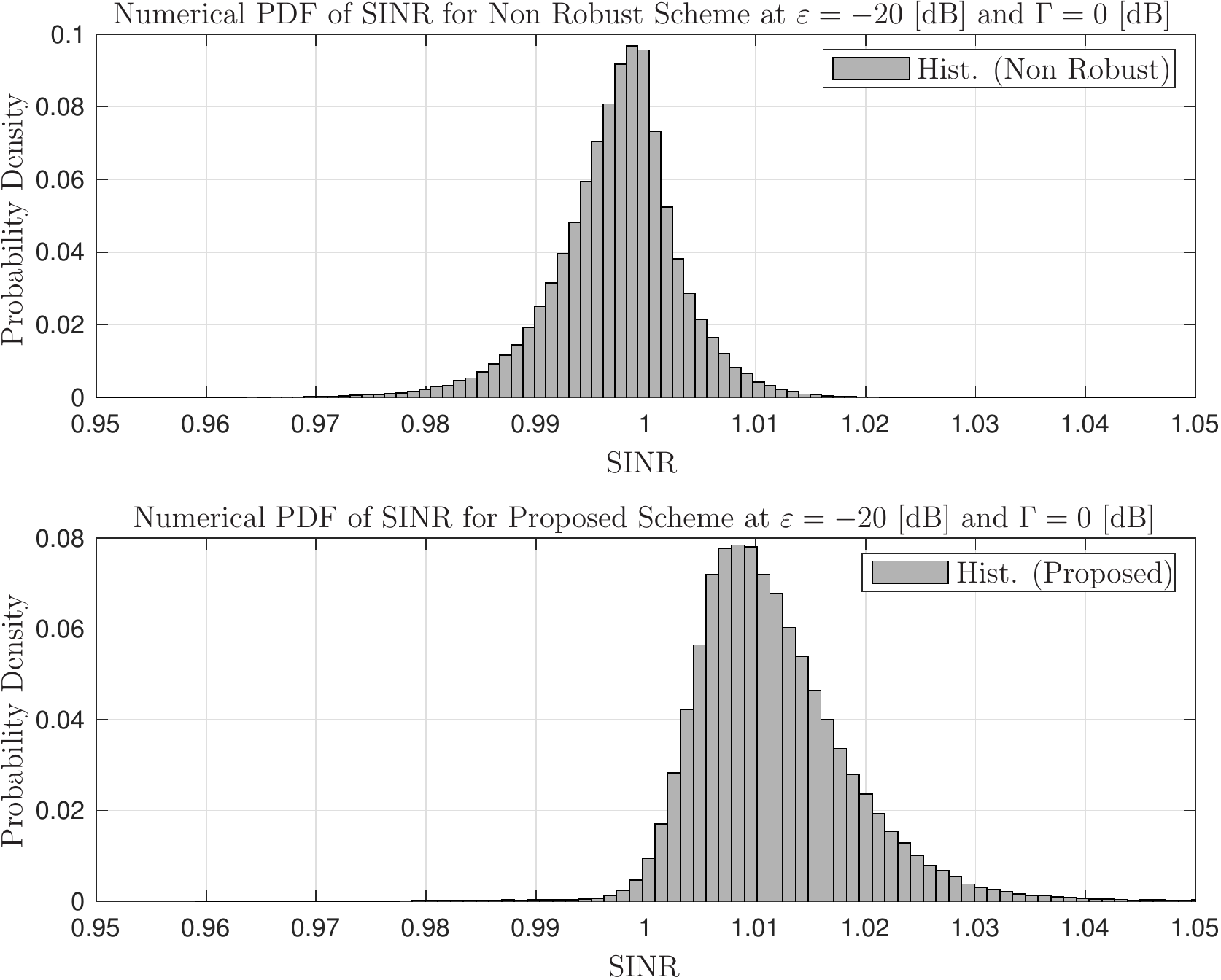}}
\subfigure[Empirical CDF.]
{\label{fig:SINRCDFat0dB}
\includegraphics[width=\columnwidth]{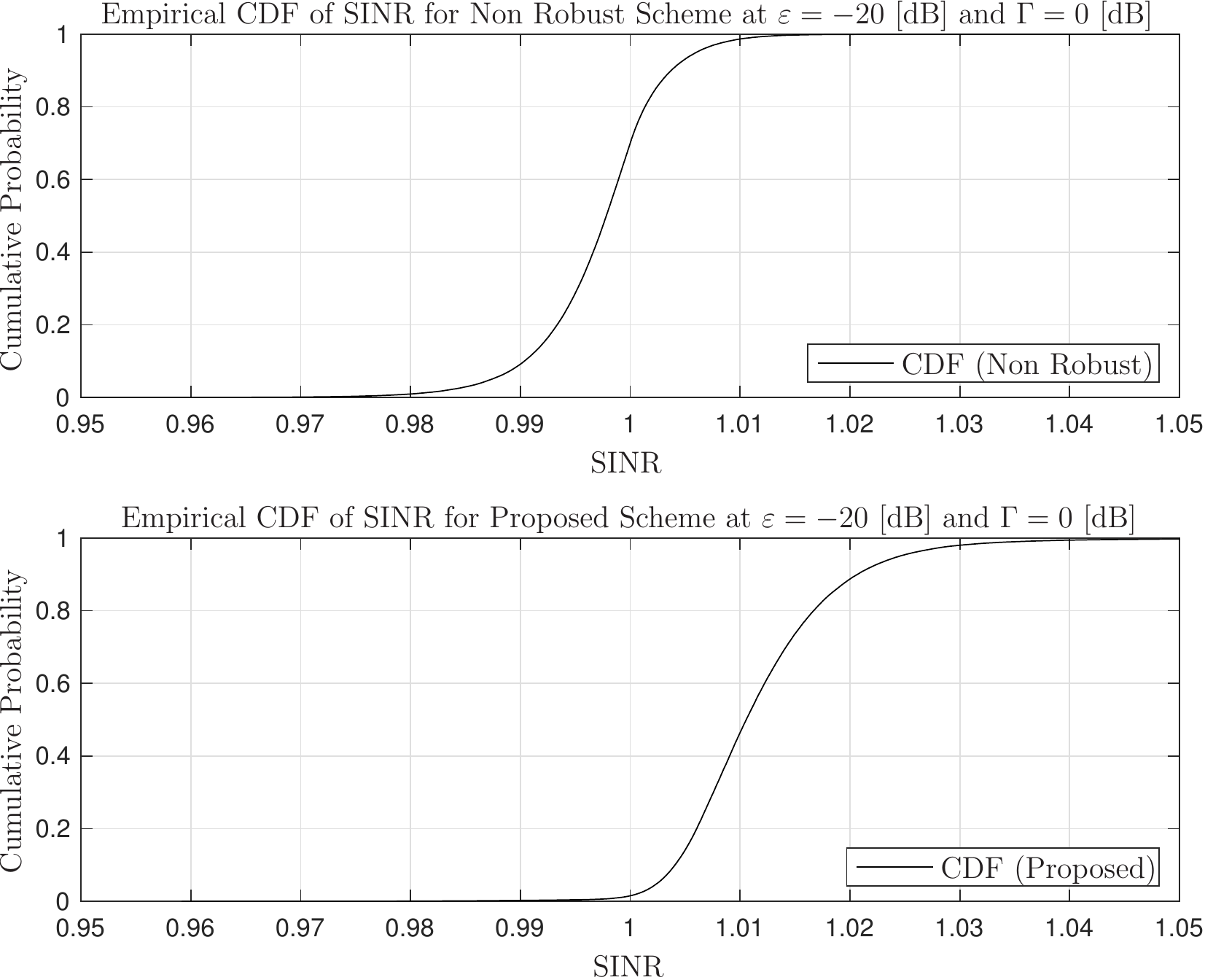}}
\caption[]{Comparisons of PDF and CDF of obtained \ac{SINR} for both the Non-Robust and the proposed schemes at $\varepsilon=0.01$, $\sigma^2=0.01$, $i_{\rm max}=10$ and $\Gamma=0$ [dB].}
\label{fig:HistCDFat0dB}
\vspace{-2ex}
\end{figure}

\begin{figure}[t!]
\centering
\subfigure[Empirical PDF.]
{\label{fig:SINRHistat10dB}
\includegraphics[width=\columnwidth]{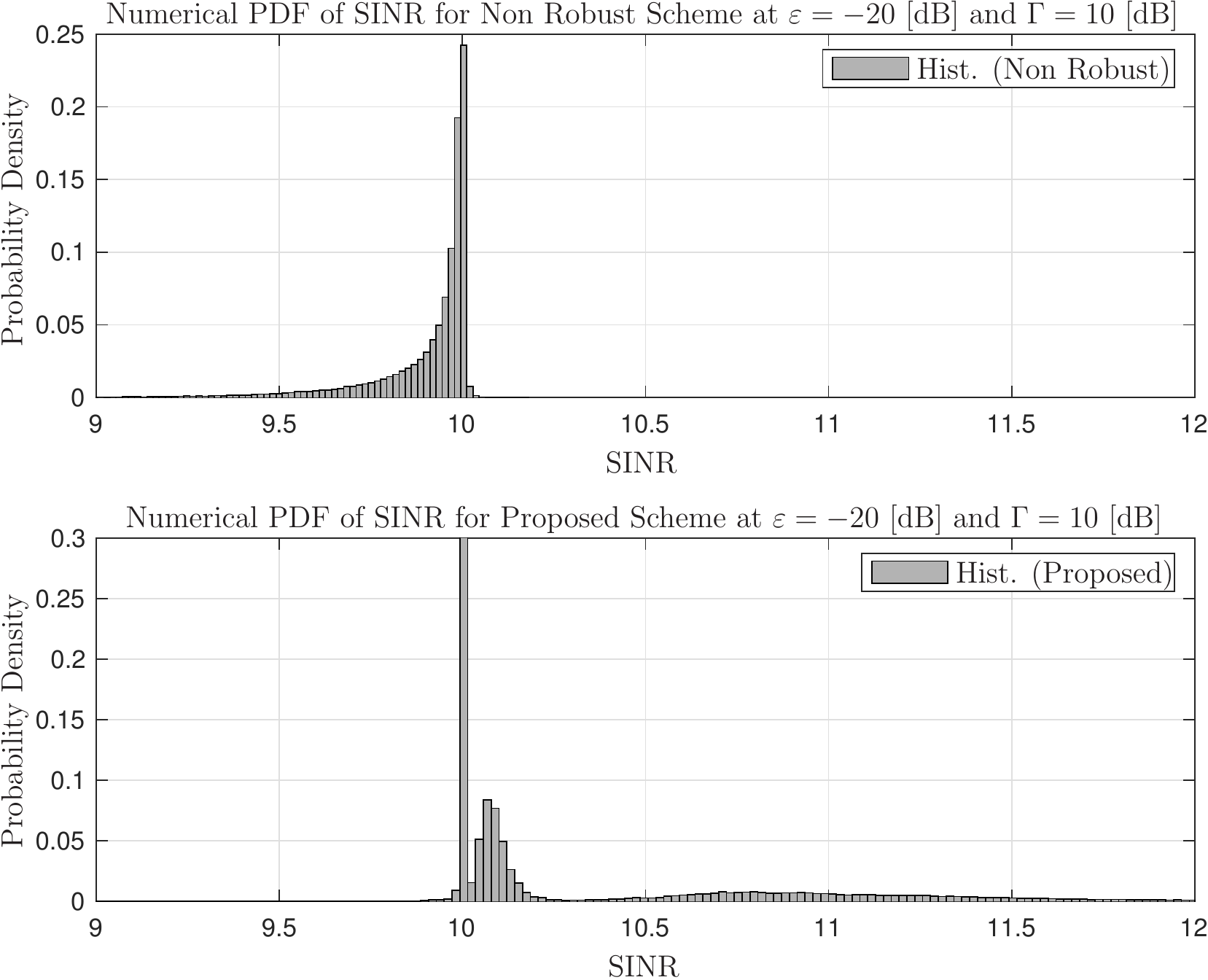}}
\subfigure[Empirical CDF]
{\label{fig:SINRCDFat10dB}
\includegraphics[width=\columnwidth]{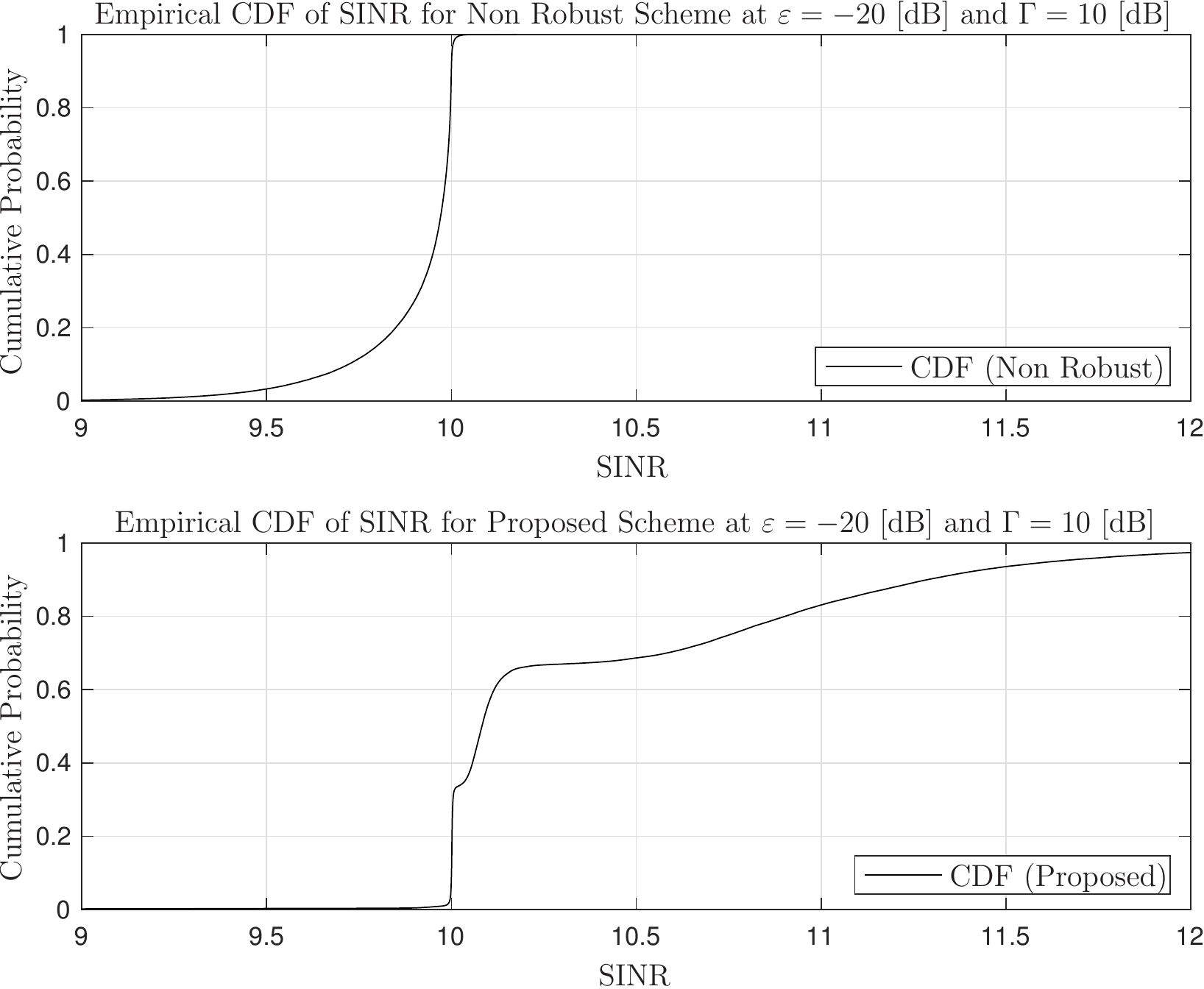}}
\caption[]{Comparisons of \ac{PDF} and \ac{CDF} of obtained \ac{SINR} for both the Non-Robust and the proposed schemes at $\varepsilon=0.01$, $\sigma^2=0.01$, $i_{\rm max}=10$ and $\Gamma=10$ [dB].}
\label{fig:HistCDFat10dB}
\end{figure}

Since this empirical distribution shown in Figure \ref{fig:IterationHist} was computed over $500$ channel realizations for each different \ac{SINR} target ranging from $0$ [dB] to $10$ [dB], it can be concluded that the convergence behavior of the proposed scheme is very stable for different \ac{QoS} requirements.
Together, Figures \ref{fig:PowervsSINRTarg} and \ref{fig:IterationHist} suggest that with the proposed robust algorithm, \ac{BB} \ac{MIMO}-\ac{NOMA} is  almost immune to the level of \ac{CSI} imperfection studied.

In Figures \ref{fig:HistCDFat0dB} and \ref{fig:HistCDFat10dB}, the empirical \ac{PDF} and corresponding \ac{CDF} of the instantaneous \acp{SINR} attained at each user over different channel and \ac{CSI} uncertainty levels are shown, for two different \ac{SINR} targets, namely, $\Gamma=0$ [dB] (in Figure \ref{fig:HistCDFat0dB}) and $\Gamma=10$ [dB] (in Figure \ref{fig:HistCDFat10dB}), respectively.
The figures further illustrate the robustness of our proposed algorithm.
In particular, it is observed that the \ac{SINR} distributions obtained as a consequence of the proposed method are significantly skewed to the right, regardless of the target \ac{QoS} requirements, indicating that negligible levels of outages are achieved.

In contrast, the corresponding distributions associated with a system designed under the assumption of perfect \ac{CSI}, but subjected to \ac{CSI} uncertainties (non-robust scheme) are found to have significant portions of their probability masses on the left side, with the condition worsening with the \ac{QoS} requirement.

Taking into account that the latter results are obtained without requiring much more power than would necessary under perfect \ac{CSI} conditions, and without requiring more than a few iterations of the algorithm, as shown in Figures \ref{fig:PowervsSINRTarg} and \ref{fig:IterationHist}, respectively, it can be confidently stated that the proposed method is effective, efficient and practical.
In addition to all the above, while generating all the data used in the figures above -- a total of $550000$ Monte-Carlo data points including $500$ channel realizations over $11$ different \ac{SINR} targets with $100$ \ac{CSI} uncertainty realizations each -- we kept track of the \emph{feasibility ration} $r_\text{Feas}$ of the proposed algorithm, defined as the fraction of times a rank-one solution of equation \eqref{eqn:OP_SDR} was obtained.
It was found that the (empirical) feasibility ratio of the method is of $r_\text{Feas}=99.99$\% at the simulated scenario of $\varepsilon=0.01$, $\sigma^2=0.01$, $N_t=3$ and $U=3$.

% Conclusion ------------------------------------
\vspace{-0.5ex}
\section{Conclusion}
\label{Sec:Conclusion}

We considered the downlink of a \ac{BB} \ac{MISO} \ac{PD-NOMA} system serving multiple single-antenna users and subjected to a norm-bounded \ac{CSI} uncertainty.
For such a system, we presented a new \ac{TX} \ac{BF} scheme, in which the total transmit power consumption is minimized subjected to prescribed \ac{SINR} requirements for each user.
The proposed scheme takes advantage of two contributions: the first is a novel  method to obtain closed-form estimates of the \ac{CSI} error vectors that result in the worst sum of the SINRs of decodable signals at each user; and the second is a \ac{FP}-based quadratic transformation of the original and intractable non-convex SINR constrained power minimization problem into a problem which is quadratic and therefore solvable using standard interior point methods.
Simulation results were shown which confirmed that the proposed algorithm ensures the robustness to imperfect \ac{CSI}, requiring only small amounts of additional transmit power compared to a system operating under idealistic (perfect \ac{CSI}) conditions.

The proposed algorithm was also shown to be fast, requiring only a few iterations to converge, and consistent, offering large gains in robust power savings even when subjected to large target SINR requirements.

\vspace{-0.5ex}
\section{Acknowledgement}
Parts of this work were supported by JSPS KAKENHI Grant Number 18H03765.

% Appendix --------------------------------------
\vspace{-1ex}
\appendices
\section{Proof of Equation \eqref{eqn:DualOP}}
\label{Appndx:A}

Although a similar problem has been discussed in \cite{OmidICC2017}, we herein provide details for the sake of complementarity.
Introducing a new slack variable $\beta_{\ell}$, the maximization of the dual function $g(\lambda_\ell)$ can be simply written in a form
\begin{subequations}
\label{eqn:AppenOP}
\begin{eqnarray}
\max_{\lambda_{\ell},\beta_{\ell}}&& \beta_{\ell}\\
\label{eqn:AppenConst1}
{\rm s.t.}&& - \mathbf{b}^{\rm H}_{\ell}\big(\lambda_{\ell}\mathbf{I}\!-\!\mathbf{A}_{\ell}\big)^{\!-1}\mathbf{b}_{\ell} + c_{\ell} \!-\! \lambda_{\ell}\varepsilon^2 \geq \beta_{\ell}\\
\label{eqn:AppenConst2}
&& \mathbf{b}_{\ell}\!\in\!\mathrm{R}(\lambda_{\ell}\mathbf{I}\!-\!\mathbf{A}_{\ell})\\
\label{eqn:AppenConst3}
&& \lambda_{\ell}\mathbf{I}-\mathbf{A}_{\ell} \succeq 0\\
&& \lambda_{\ell} \geq 0.
\end{eqnarray}
\end{subequations}

Recalling the Schur complement, we obtain
\begin{eqnarray}
\mathbf{X}\succeq 0 &\Leftrightarrow& 
  \begin{cases}
\lambda_{\ell}\mathbf{I}-\mathbf{A}_{\ell}\succeq 0\:\: \nonumber\\
\mathbf{b}_{\ell}\!\in\!\mathrm{R}(\lambda_{\ell}\mathbf{I}\!-\!\mathbf{A}_{\ell})\\
- \mathbf{b}^{\rm H}_{\ell}\big(\lambda_{\ell}\mathbf{I}\!-\!\mathbf{A}_{\ell}\big)^{\!-1}\mathbf{b}_{\ell} + c_{\ell} \!-\! \lambda_{\ell}\varepsilon^2 - \beta_{\ell}\geq 0\nonumber
  \end{cases}
\end{eqnarray}
where
\begin{equation}
\mathbf{X} = \begin{bmatrix}
\lambda_{\ell}\mathbf{I}\!-\!\mathbf{A}_{\ell} & \mathbf{b}_{\ell}\\
\mathbf{b}^{\rm H}_{\ell} & c_{\ell} \!-\! \lambda_{\ell}\varepsilon^2 - \beta_{\ell}
\end{bmatrix}.
\end{equation}
\begin{proof}
See Proposition 2.1 and its proof in \cite{TingCOA2014}.
For more details about the (generalized) trust region problem, see \cite{TingCOA2014} and references thereby.
\end{proof}

\bibliographystyle{IEEEtran}
% Ignore errors thrown by references section before editing
\bibliography{IEEEabrv,\myreferences,MyListOfPapers}

\begin{thebibliography}{10}
\providecommand{\url}[1]{#1}
\csname url@samestyle\endcsname
\providecommand{\newblock}{\relax}
\providecommand{\bibinfo}[2]{#2}
\providecommand{\BIBentrySTDinterwordspacing}{\spaceskip=0pt\relax}
\providecommand{\BIBentryALTinterwordstretchfactor}{4}
\providecommand{\BIBentryALTinterwordspacing}{\spaceskip=\fontdimen2\font plus
\BIBentryALTinterwordstretchfactor\fontdimen3\font minus
  \fontdimen4\font\relax}
\providecommand{\BIBforeignlanguage}[2]{{%
\expandafter\ifx\csname l@#1\endcsname\relax
\typeout{** WARNING: IEEEtran.bst: No hyphenation pattern has been}%
\typeout{** loaded for the language `#1'. Using the pattern for}%
\typeout{** the default language instead.}%
\else
\language=\csname l@#1\endcsname
\fi
#2}}
\providecommand{\BIBdecl}{\relax}
\BIBdecl

\bibitem{Glisic1997}
S.~G. Glisic and P.~A. Lepp\"{a}nen, \emph{Wireless Communications: {TDMA}
  versus {CDMA}}.\hskip 1em plus 0.5em minus 0.4em\relax Springer, 1997.

\bibitem{Viterbi1995}
A.~J. Viterbi, \emph{{CDMA}: Principles of Spread Spectrum
  Communication}.\hskip 1em plus 0.5em minus 0.4em\relax Redwood City, USA:
  Addison Wesley Longman Publishing Co., Inc., 1995.

\bibitem{Holma2009}
H.~Holma and A.~Toskala, \emph{{LTE} for {UMTS} - {OFDMA} and {SC-FDMA} Based
  Radio Access}.\hskip 1em plus 0.5em minus 0.4em\relax Wiley, 2009.

\bibitem{Ding2017}
Z.~Ding, X.~Lei, G.~K. Karagiannidis, R.~Schober, J.~Yuan, and V.~K. Bhargava,
  ``A survey on non-orthogonal multiple access for {5G} networks: Research
  challenges and future trends,'' \emph{{IEEE J. Sel. Areas Commun.}}, vol.~35,
  no.~10, pp. 2181--2195, Oct. 2017.

\bibitem{Saito2013}
Y.~Saito, Y.~Kishiyama, A.~Benjebbour, T.~Nakamura, A.~Li, and K.~Higuchi,
  ``{Non-Orthogonal Multiple Access (NOMA) for Cellular Future Radio Access},''
  in \emph{Proc. {IEEE} VTC Spring}, Dresden, Germany, Jun. 2013, pp. 1--5.

\bibitem{LiuPIEEE2017}
Y.~Liu, Z.~Qin, M.~Elkashlan, Z.~Ding, A.~Nallanathan, and L.~Hanzo,
  ``Nonorthogonal multiple access for {5G} and beyond,'' \emph{{Proceedings of
  the IEEE}}, vol. 105, no.~12, pp. 2347--2381, Dec. 2017.

\bibitem{DingTWC2016}
Z.~Ding, F.~Adachi, and H.~V. Poor, ``The application of {MIMO} to
  non-orthogonal multiple access,'' \emph{{IEEE Trans. Wireless Commun.}},
  vol.~15, no.~1, pp. 537--552, Jan. 2016.

\bibitem{Elkashlan2016}
Y.~L. .~M. Elkashlan, Z.~Ding, and G.~K. Karagiannidis, ``Fairness of user
  clustering in {MIMO} non-orthogonal multiple access systems,'' \emph{{IEEE
  Commun. Letters}}, vol.~20, no.~7, pp. 1465--1468, Jul. 2016.

\bibitem{AliAccess2016}
M.~S. Ali, E.~Hossain, and D.~I. Kim, ``Non-orthogonal multiple access ({NOMA})
  for downlink multiuser {MIMO} systems: User clustering, beamforming, and
  power allocation,'' \emph{{IEEE Access}}, vol.~5, pp. 565--577, Dec. 2016.

\bibitem{HanifTSP2016}
M.~F. Hanif, Z.~Ding, T.~Ratnarajah, and G.~K. Karagiannidis, ``A
  minorization-maximization method for optimizing sum rate in the downlink of
  non-orthogonal multiple access systems,'' \emph{{IEEE Trans. Signal
  Process.}}, vol.~64, no.~1, pp. 76--88, Jan. 2016.

\bibitem{ChoiTWC2016}
J.~Choi, ``On the power allocation for {MIMO-NOMA} systems with layered
  transmissions,'' \emph{{IEEE Trans. Wireless Commun.}}, vol.~15, no.~5, pp.
  3226--3237, May 2016.

\bibitem{AlaviTVT2018}
F.~Alavi, K.~Cumanan, Z.~Ding, and A.~G. Burr, ``Beamforming techniques for
  nonorthogonal multiple access in {5G} cellular networks,'' \emph{{IEEE Trans.
  Veh. Technol.}}, vol.~67, no.~10, pp. 9474--9487, Oct. 2018.

\bibitem{MyListOfPapers:StanzakTSP2006}
S.~Sta{\'n}zak, G.~Wunder, and H.~Boche, ``On pilot-based multipath channel
  estimation for uplink {CDMA} systems: An overloaded case,'' \emph{{IEEE
  Trans. Signal Process.}}, vol.~54, no.~2, pp. 512 -- 519, Feb. 2006.

\bibitem{AlaviCL2017}
F.~Alavi, K.~Cumanan, Z.~Ding, and A.~G. Burr, ``Robust beamforming techniques
  for non-orthogonal multiple access systems with bounded channel
  uncertainties,'' \emph{{IEEE Commun. Letters}}, vol.~21, no.~9, pp.
  2033--2036, Sep. 2017.

\bibitem{Bjornson2013}
E.~Bj\"{o}rnson and E.~Jorswieck, ``Optimal resource allocation in coordinated
  multi-cell systems,'' \emph{{Foundations and Trends in Communications and
  Information Theory}}, vol.~9, no. 2--3, pp. 113--381, Jan. 2013.

\bibitem{ShenTSP2018}
K.~Shen and W.~Yu, ``Fractional programming for communication systems -- {P}art
  {I}: {P}ower control and beamforming,'' \emph{{IEEE Trans. Signal Process.}},
  vol.~66, no.~10, pp. 2616--2630, May 2018.

\bibitem{IimoriTWC19}
H.~Iimori, G.~T.~F. de~Abreu, and G.~C. Alexandropoulos, ``{MIMO} beamforming
  schemes for hybrid {SIC FD} radios with imperfect hardware and {CSI},''
  \emph{{IEEE Trans. Wireless Commun.}}, vol.~18, no.~10, pp. 4816--4830, Oct.
  2019.

\bibitem{IimoriWCNC19}
H.~Iimori, G.~Abreu, K.~Ishibashi, and G.~C. Alexandropoulos, ``Transmission
  strategies in imperfect bi-directional full-duplex {MIMO} systems,'' in
  \emph{{Proc. IEEE WCNC}}, Marrakesh,Morocco, Apr. 2019.

\bibitem{Zheng2012}
X.~Zheng, X.~Sun, D.~Li, and Y.~Xu, ``On zero duality gap in nonconvex
  quadratic programming problems,'' \emph{J. Glob. Optim.}, vol.~52, no.~2, pp.
  229--242, Feb. 2012.

\bibitem{Pong2014}
T.~Pong and H.~Wolkowicz, ``The generalized trust region subproblem,''
  \emph{Comput. Optim. Appl.}, vol.~58, no.~2, pp. 273--322, Jan. 2014.

\bibitem{IimoriSPAWC2018}
H.~Iimori and G.~Abreu, ``Two-way full-duplex {MIMO} with hybrid {TX-RX} {MSE}
  minimization and interference cancellation,'' in \emph{{Proc. IEEE SPAWC}},
  Kalamata, Greece, 25-28 Jun. 2018, pp. 1--6.

\bibitem{CVX2017}
\BIBentryALTinterwordspacing
M.~Grant and S.~Boyd. (2017) {CVX}: Matlab software for disciplined convex
  programming. [Online]. Available: \url{http://cvxr.com/cvx/}
\BIBentrySTDinterwordspacing

\bibitem{AlexSPM2010}
A.~B. Gershman, N.~D. Sidiropoulos, S.~Shahbazpanahi, M.~Bengtsson, and
  B.~Ottersten, ``Convex optimization-based beamforming,'' \emph{{IEEE Signal
  Process. Mag.}}, vol.~27, no.~3, pp. 62--75, Apr. 2010.

\bibitem{KaripidisTSP08}
E.~Karipidis, N.~D. Sidiropoulos, and Z.-Q. Luo, ``Quality of service and
  max-min fair transmit beamforming to multiple cochannel multicast groups,''
  \emph{{IEEE Trans. Signal Process.}}, vol.~56, no.~3, pp. 1268--1279, Mar.
  2008.

\bibitem{LuoSPM10}
Z.~quan Luo, W.~kin Ma, A.~M. cho So, Y.~Ye, and S.~Zhang, ``Semidefinite
  relaxation of quadratic optimization problems,'' \emph{{IEEE Signal Process.
  Mag.}}, vol.~27, no.~3, pp. 20--34, May 2010.

\bibitem{OmidICC2017}
O.~Taghizadeh and R.~Mathar, ``Worst-case robust sum rate maximization for
  full-duplex bi-directional {MIMO} systems under channel knowledge
  uncertainty,'' in \emph{{Proc. IEEE ICC}}, Paris, France, May 2017, pp. 1--7.

\bibitem{TingCOA2014}
T.~K. Pong and H.~Wolkowicz, ``The generalized trust region subproblem,''
  \emph{Comput. Optim. Appl.}, vol.~58, no.~2, pp. 273--322, Jan. 2014.

\end{thebibliography}
\end{document}